\documentclass[aps,prl,twocolumn,superscriptaddress,longbibliography]{revtex4-2}  %
\usepackage{graphicx} 
\usepackage{float} 
\usepackage{dcolumn} 
\usepackage{bm,color}
\usepackage{amssymb}
\usepackage{amsmath}
\usepackage{extarrows}
\usepackage[colorlinks=true,linkcolor=blue,urlcolor=blue,citecolor=blue]{hyperref}

\usepackage[normalem]{ulem} 
\definecolor{mgreen}{RGB}{1,123,0}

\begin{document}
\title{Anyonization of bosons in one dimension: an effective swap model}

\author{ Botao Wang}
\email[]{botao.wang@ulb.be}
\affiliation{Center for Nonlinear Phenomena and Complex Systems, Université Libre de Bruxelles, CP 231, Campus Plaine, 1050 Brussels, Belgium}
\affiliation{International Solvay Institutes, 1050 Brussels, Belgium}

\author{ Amit Vashisht }
\email[]{amit.vashisht@ulb.be}
\affiliation{Center for Nonlinear Phenomena and Complex Systems, Université Libre de Bruxelles, CP 231, Campus Plaine, 1050 Brussels, Belgium}
\affiliation{International Solvay Institutes, 1050 Brussels, Belgium}

\author{ Yanliang  Guo }
\email{yanliang.guo@uibk.ac.at}
\affiliation{Institut f{\"u}r Experimentalphysik und Zentrum f{\"u}r Quantenphysik, Universit{\"a}t Innsbruck, Technikerstra{\ss}e 25, Innsbruck, 6020, Austria}

\author{ Sudipta  Dhar }
\affiliation{Institut f{\"u}r Experimentalphysik und Zentrum f{\"u}r Quantenphysik, Universit{\"a}t Innsbruck, Technikerstra{\ss}e 25, Innsbruck, 6020, Austria} 

\author{ Manuele  Landini}
\affiliation{Institut f{\"u}r Experimentalphysik und Zentrum f{\"u}r Quantenphysik, Universit{\"a}t Innsbruck, Technikerstra{\ss}e 25, Innsbruck, 6020, Austria}

\author{ Hanns-Christoph  N{\"a}gerl}
\affiliation{Institut f{\"u}r Experimentalphysik und Zentrum f{\"u}r Quantenphysik, Universit{\"a}t Innsbruck, Technikerstra{\ss}e 25, Innsbruck, 6020, Austria}

\author{Nathan Goldman}
\email[]{nathan.goldman@ulb.be}
\affiliation{Center for Nonlinear Phenomena and Complex Systems, Université Libre de Bruxelles, CP 231, Campus Plaine, 1050 Brussels, Belgium}
\affiliation{International Solvay Institutes, 1050 Brussels, Belgium}
\affiliation{Laboratoire Kastler Brossel, Collège de France, CNRS, ENS-Université PSL, Sorbonne Université, 11 Place Marcelin Berthelot, 75005 Paris, France}

\date{\today}

\begin{abstract}
    Anyons emerge as elementary excitations in low-dimensional quantum systems and exhibit behavior distinct from bosons or fermions. Previous models of anyons in one dimension (1D) are mainly categorized into two types: those that rely on nontrivial scattering behavior, and those based on density-dependent hopping processes in discrete lattices. Here, we introduce a novel framework for realizing anyonic correlations using the internal degrees of freedom of a spinor quantum gas. We propose a ``swap" model, which assigns a complex phase factor to the swapping processes between two different species, referred to as ``host particles'' and ``impurities". The anyonic characteristics are demonstrated through the one-body correlator of the impurity, using a spin-charge separation analysis. For a single impurity, our swap model can be effectively implemented by applying tilt potentials in a strongly interacting quantum gas [Dhar \textit{et al.}, arXiv:2412.21131]. We further explore the dynamical properties of anyonic correlations and extend our analysis to the case of multiple impurities. Our work provides new avenues for engineering many-body anyonic behavior in quantum simulation platforms.
\end{abstract}

\maketitle

\paragraph{Introduction.}
Exotic behaviour can emerge in lower-dimensional systems. A prime example is given by the emergence of anyons obeying fractional statistics, interpolating between bosons and fermions~\cite{1977Leinaas,1981Goldin,1982Wilczek,2005Khare,2024Greiter}.
Arising from a mathematical concept, anyons are often found to exist as elementary excitations of topologically-ordered states of matter~\cite{1982Tsui,1983Laughlin,1984Halperin,1984Arovas,2021Feldman,2006Kitaev_anyons,2007Yao_exact,2014Bauer}.
The idea of utilizing non-Abelian anyons for topological quantum computing has spurred efforts to observe their fractional statistics across a variety of physical platforms, such as solid-state materials~\cite{2020Bartolomei,2020Nakamura}, superconducting quantum circuits~\cite{2016Zhong,2021Satzinger,2023Google,2024Xu}, Rydberg atom arrays~\cite{2021Semeghini} and trapped-ion processors~\cite{2024Iqbal}.
Since anyons are typically defined through the exchange of quasiparticles in two dimensions (2D) with topologically distinct windings, fractional statistics seems to be absent in one dimension (1D).

The generalization of fractional statistics to arbitrary dimensions was provided by Haldane~\cite{1991Haldane}. He demonstrated that the spinon excitations in a 1D spin chain, described by the Haldane-Shastry model (HSM)~\cite{1988Haldane,1988Shastry}, satisfy fractional exclusion statistics~\cite{1991Haldane}. Subsequent investigations predicted the presence of 1D (Abelian) anyonic excitations in other models, including the Kuramoto-Yokoyama model~\cite{1991Kuramoto,1995Kuramoto,1998Kato,2001Arikawa} and the Calogero-Sutherland model~\cite{1994Ha,1995Ha,1994Murthy}. 
It was recognized that these excitations exhibit a common feature: a statistical phase can be associated to their unidirectional crossing, leading to a fractional shift in the linear momenta in 1D that is analogous to the fractional relative angular momentum in 2D~\cite{2009Greiter,2024Greiter}.
By leveraging the notion of statistical interactions~\cite{1991Haldane}, one can embed fractional statistics into 1D scattering processes, offering an interesting perspective to reinterpret interacting bosons~\cite{1994Isakov,1994Wu,1995Wu,2006Batchelor,1999Kundu,1992Hansson,2017Posske}.
Experimentally, realizing and probing anyons in such 1D models has been much less explored. Recently, the chiral BF theory~\cite{1996Aglietti,1999Kundu,2022Chisholm}, originally constructed as a candidate model for 1D anyons, has been realized in a Raman-coupled Bose-Einstein condensate~\cite{2022Chisholm,2022Frolian}.

Another route to realize 1D Abelian anyons is offered by engineering density-dependent Peierls phases on a lattice, following the bosonic formulation of the anyon-Hubbard model (AHM)~\cite{2011Keilmann,2015Greschner,2016Cardarelli,2016Straeter,2017Yuan}. This approach has been shown to give rise to intriguing phenomena, such as statistically driven equilibrium phases~\cite{2016Forero,2018Forero,2017Lange,2022Olekhno,2024Bonkhoff} and non-equilibrium anyonic dynamics~\cite{2012Hao,2014Wang,2014Wright,2018Liu,2018Greschner,zhang_observation_2022,2022Wang_exact}.
Despite extensive theoretical efforts~\cite{2011Keilmann,2015Greschner,2016Cardarelli,2016Straeter,2017Yuan,2016Forero,2018Forero,2017Lange,2022Olekhno,2024Bonkhoff,2012Hao,2014Wang,2014Wright,2018Liu,2018Greschner,zhang_observation_2022,2022Wang_exact,2006Girardeau,2021Bonkhoff,wang2024boson}, experimental realization of 1D anyons was only recently observed in an optical lattice using two ultracold atoms, where the density-dependent hopping was engineered using Floquet driving~\cite{2024Kwan}.
However, a direct measurement of the asymmetric momentum distribution, the hallmark signature of anyonic correlations~\cite{Calabrese2007,Patu2007,Santachiara2007,2008Santachiara,2008Hao,2009Hao,2013Mintchev,2020Scopa}, is not possible in that setting:\ the main reason is that anyonic correlations need to be manually reconstructed upon extracting the momentum distribution of the bare bosons.

More recently, anyonic correlations were observed by injecting an impurity into a strongly-interacting Bose gas~\cite{2024Dhar}, wherein the statistical phase was encoded in the momentum of the impurity, hence allowing for a direct measurement of the asymmetric momentum distribution~\cite{2017Yang,2020Gamayun,2024Gamayun}. Here, we present a simple yet insightful model to capture the essence of anyonization in a strongly interacting 1D spinor quantum gas, providing a general framework to engineer anyonic correlations in 1D. 
We propose a new ``swap'' model, which associates the exchange between the two species of the spinor quantum gas with a tunable complex phase. Using a spin-charge separation analysis, we demonstrate anyonic signatures in the impurity's one-body correlation function. For a single impurity, this scheme can be implemented by generating a spin wave, for instance, by applying a weak constant force to the impurity. This simple picture offers an effective description of the recent experimental observations of Ref.~\cite{2024Dhar}. We then extend our approach to study the real-time evolution of anyonic correlations as well as scenarios with multiple impurities. Altogether, this work lays the groundwork for engineering and probing anyonic many-body behaviour in 1D quantum simulators.

\begin{figure}
	\centering\includegraphics[width=0.98\linewidth]{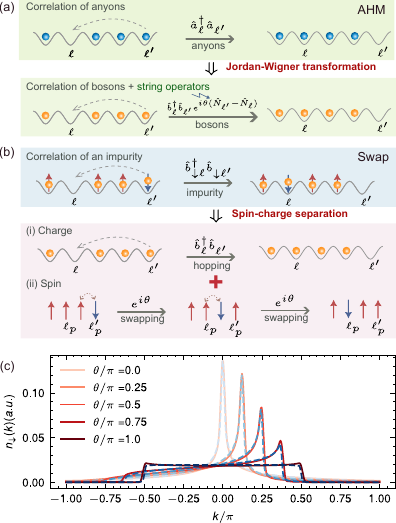} %
	\caption{(a) Sketch of one-body correlations $\langle\hat{a}_{\ell}^{\dagger}\hat{a}_{\ell'}\rangle$ of anyons in the AHM (w.r.t. a given Fock state configuration). By means of a Jordan-Wigner transformation, the anyonic one-body correlator can be expressed as that of bosons accompanied by a phase factor, i.e.\ $\langle\hat{a}_{\ell}^{\dagger}\hat{a}_{\ell'}\rangle\!=\!\langle\hat{b}_{\ell}^{\dagger}e^{i\theta(\hat{N}_{\ell'}-\hat{N}_{\ell})}\hat{b}_{\ell'}\rangle$. (b) Sketch of one-body correlation $\langle\hat{b}_{\downarrow\ell}^{\dagger}\hat{b}_{\downarrow\ell'}\rangle$ of a single impurity in our swap model. As there are two host particles between site $\ell$ and $\ell'$, the long-range hopping indicated here involves two swaps. Since each swap is associated with $e^{i\theta}$, the swapping processes contribute a phase factor $e^{i\theta(\ell'_p-\ell_p)}\!=\!e^{i2\theta}$ that recovers the contribution from the string operators in (a). (c) Quasi-momentum distribution of the impurity with different phase $\theta$. Solid lines: Using our swap Hamiltonian~(\ref{H_swap}) with $N_\uparrow\!=\!50, N_\downarrow=1,  L\!=\!100, J_{\text{ex}}\!=\!0.1$. Dashed lines: results of AHM with $N\!=\!50, L\!=\!100$ and hard-core interaction. The peak has been rescaled according to a normalization factor determined at $\theta=0$.
    } 
	\label{fig1}
\end{figure}

\paragraph{Anyon-Hubbard model.}
We start by briefly reviewing the properties of anyons in a 1D lattice, as described by the paradigmatic AHM,
\begin{equation}
	\hat{H}_{\text{AHM}}=-J\sum_{\ell}\left(\hat{a}_{\ell}^{\dagger}\hat{a}_{\ell+1}+h.c.\right)+\frac{U}{2}\sum_{\ell}\hat{n}_{\ell}\left(\hat{n}_{\ell}-1\right),
	\label{H_AHM}
\end{equation}
where $J$ denotes the tunneling amplitude; $U$ quantifies the on-site interaction strength between anyons and $\hat{n}_{\ell}\!=\!\hat{a}_{\ell}^{\dagger}\hat{a}_{\ell} $ is the number operator at site $\ell$. The anyonic operators $\hat{a}_\ell$ obey the generalized commutation relations,
\begin{subequations}
\begin{align}
  \hat{a}_{j}^{\dagger}\hat{a}_{k}-e^{-i\theta\text{sgn}(j-k)}\hat{a}_{k}^{\dagger}\hat{a}_{j} &= \delta_{jk} \\
  \hat{a}_{j}\hat{a}_{k}-e^{-i\theta\text{sgn}(j-k)}\hat{a}_{k}\hat{a}_{j} &= 0,
\end{align}
\end{subequations}
where $\theta$ is the statistical angle.
The sign function $\text{sgn}(j-k)$ equals $+1$ for $j > k$, $-1$ for $j < k$ and $0$ for $j=k$.
By applying a fractional version of the Jordan-Wigner (JW) transformation, $\hat{a}_{\ell}\!=\!\hat{b}_{\ell}e^{i\theta \hat{N}_{\ell}}$, where $\hat{N}_{\ell}\!=\!\sum_{m<\ell}\hat{n}_{m}$ and $\hat{b}_\ell$ are bosonic operators,
the one-body correlation operator reads $\hat{a}_{\ell}^{\dagger}\hat{a}_{\ell'}\!=\!\hat{b}_{\ell}^{\dagger}e^{i\theta(\hat{N}_{\ell'}-\hat{N}_{\ell})}\hat{b}_{\ell'}$. Writing the ground state in the Fock state basis $|\Psi\rangle\!=\!\sum_{i}c_{i}|\zeta_{i}\rangle$, with $c_i$ being complex coefficients, one can express the anyonic correlation function as
\begin{equation}
	\langle\hat{a}_{\ell}^{\dagger}\hat{a}_{\ell'}\rangle=\begin{cases}
		{\displaystyle \sum_{i,j}}c_{i}^{*}c_{j}\langle\zeta_{i}|\hat{b}_{\ell}^{\dagger}\hat{b}_{\ell'}|\zeta_{j}\rangle e^{i\theta(N_{\ell'}^{j}-N_{\ell}^{j})}, & \ell'\geq\ell,\\
		{\displaystyle \sum_{i,j}}c_{i}^{*}c_{j}\langle\zeta_{i}|\hat{b}_{\ell}^{\dagger}\hat{b}_{\ell'}|\zeta_{j}\rangle e^{i\theta(N_{\ell'}^{j}-N_{\ell}^{j}+1)}, & \ell'<\ell,
	\end{cases}
	\label{corr_ab_fock}
\end{equation}
where $N_{\ell}^{j}$ denotes the number of particles after applying the operator $\hat{N}_{\ell}$ on the Fock state $|\zeta_{j}\rangle$, and $\langle \ldots \rangle$ denotes an expectation value over the ground state of the system.
As shown in Eq.(\ref{corr_ab_fock}), the string operators entering the JW transformation lead to a phase factor depending on the density difference in the one-body correlation [see Fig.~\ref{fig1}(a)].
By solving the time-independent Schr\"odinger equation associated with the AHM in the bosonic representation, and performing the Fourier transformation of the one-body correlation function (\ref{corr_ab_fock}), one obtains the $\theta$-dependent asymmetric quasi-momentum distribution of the anyons~\cite{2009Hao,2015Tang}; see the dashed lines in Fig.~\ref{fig1}(c). This asymmetric behavior stands as a hallmark signature of anyons in 1D systems.

\paragraph{Swap model.}
Previous 1D anyon models often include the statistical angle either in the on-site interaction terms~\cite{1994Isakov,1994Wu,1995Wu,1999Kundu,2006Batchelor,1992Hansson,2017Posske} or in the density-dependent hopping terms~\cite{2011Keilmann,2015Greschner,2016Cardarelli,2016Straeter}. Here, instead, we propose a two-component model that encodes the anyonic correlations using an internal (`spin') degree of freedom. Considering a strongly-interacting spinor quantum gas in a 1D lattice, the proposed swap model reads
\begin{align}
    \hat{H}_{\text{swap}}= & -J{\displaystyle \sum_{\ell}}\left(\hat{b}_{\uparrow\ell}^{\dagger}\hat{b}_{\uparrow\ell+1}+\hat{b}_{\downarrow\ell}^{\dagger}\hat{b}_{\downarrow\ell+1}\right)
    \nonumber \\ &
    -J_{\text{ex}}e^{i\theta}\sum_{\ell=1}^{L-1}\hat{b}_{\uparrow\ell}^{\dagger}\hat{b}_{\downarrow\ell+1}^{\dagger}\hat{b}_{\downarrow\ell}\hat{b}_{\uparrow\ell+1}+\text{h.c.},
    \label{H_swap}
\end{align}
where $\hat{b}^\dagger_{\sigma\ell}$ ($\hat{b}_{\sigma\ell}$) are the creation (annihilation) operators of hardcore bosons with spin $\sigma\!=\!\{\uparrow,\downarrow\}$ at site $\ell$. We use spin indices $\uparrow$ and $\downarrow$ to denote the host and the impurity particles, respectively; $J$ represents the strength of nearest-neighbor hopping and $J_{\rm ex}$ is the strength of the swapping between different spins on neighboring sites. Here we assume no double occupancy, and thus our model applies in the regime of strong on-site interaction. This is reminiscent of the $tJ$-model studied in quantum magnetism~\cite{2019Eckle}. One major difference is the complex phase factor $e^{i\theta}$ assigned to the swapping terms. In the following, we will show that $\theta$ plays the role of the fractional statistical angle of anyons, as captured by the one-body correlation function. We use $L$ as the number of lattice sites and $N_{\uparrow} (N_{\downarrow})$ as the number of host particles (impurities). In the following, we first focus on the case of a single impurity.

We compute the one-body correlation of the impurity based on a spin-charge separation analysis~\cite{2015Yang,2023Basak}. 
In the Fock state basis $\{|\psi_{i}\rangle\}$, the one-body correlation function of the impurity reads 
\begin{equation}
	\langle\hat{b}_{\downarrow\ell}^{\dagger}\hat{b}_{\downarrow\ell'}\rangle={\displaystyle \sum_{i,j}}\tilde{c}_{i}^{*}\tilde{c}_{j}\langle\psi_{i}|\hat{b}_{\downarrow\ell}^{\dagger}\hat{b}_{\downarrow\ell'}|\psi_{j}\rangle,
\end{equation}
with $\tilde{c}_{i}$ being complex coefficients. 
Each Fock state can be written as the product of its charge sector $|\psi_{i}^{c}\rangle$ and the spin sector $|\psi_{i}^{s}\rangle$, i.e. $|\psi_{i}\rangle\!=\!|\psi_{i}^{c}\rangle\otimes|\psi_{i}^{s}\rangle$.
Meanwhile, we can decompose the one-body correlation operator of the impurity $\hat{b}_{\downarrow\ell}^{\dagger}\hat{b}_{\downarrow\ell'}$ as a hopping of the charge and a swapping of the spins. 
Taking the Fock state configuration sketched in Fig.~\ref{fig1}(b) as a concrete example, the one-body correlator consists of the hopping of the charge from site $\ell'$ to $\ell$, combined with a series of swapping processes that start from the spin labeled by $\ell'_p\!=\!N_{\ell'}^{j}+1\!=\!4$ and end with the spin $\ell_p\!=\!N_{\ell}^{j}+1\!=\!2$. Here, we used $\ell$ and $\ell_p$ to label the lattice sites and the spins, respectively, and $N_{\ell}^{j}$ denotes the number of particles to the left of site $\ell$ in a given Fock state $|\psi^c_{j}\rangle$.  It is apparent that the number of swaps is determined by the number of particles between site $\ell$ and $\ell'$, i.e.\ $\ell'_{p}-\ell_{p}\!=\!N_{\ell'}^{j}-N_{\ell}^{j}$. 

Based on this observation, one can write the correlator in the given Fock states as
\begin{align}
\langle\psi_{i}|\hat{b}_{\downarrow\ell}^{\dagger}\hat{b}_{\downarrow\ell'}|\psi_{j}\rangle= & \langle\psi_{i}^{c}|\hat{b}_{\ell}^{\dagger}\hat{b}_{\ell'}|\psi_{j}^{c}\rangle\delta_{N_{\ell}^{j}}^{\ell_{p}-1}\delta_{N_{\ell'}^{j}}^{\ell_{p}'-1}
\\ \nonumber
& \times\langle\psi_{i}^{s}|\hat{\mathcal{E}}_{\ell_{p},\ell_{p}\mp1}\cdots\hat{\mathcal{E}}_{\ell'_{p}\pm1,\ell'_{p}}|\psi_{j}^{s}\rangle.
\end{align}
Here, the Kronecker delta, $\delta_{N_{\ell}^{j}}^{\ell_{p}-1}\!=\!1$ if $\ell_{p}-1\!=\!N_{\ell}^{j}$ and 0 otherwise, ensures that the $\ell'$-th site is occupied by the $\ell'_p$-th spin, and after hopping, the $\ell$-th site is occupied by the $\ell_p$-th spin~\cite{2023Basak}. 
We use $\hat{\mathcal{E}}_{\ell_{p}\pm1,\ell_{p}}$ to denote the swapping between the impurity labeled by $\ell_p$ and the host particle labeled by $\ell_p\pm1$.
Assigning a phase factor $e^{\pm i\theta}$ in the swapping between the impurity and the host particle to its left (right), we obtain the following one-body correlation function of the impurity
\begin{equation}
\langle\hat{b}_{\downarrow\ell}^{\dagger}\hat{b}_{\downarrow\ell'}\rangle=\begin{cases}
{\displaystyle \sum_{i,j}\tilde{c}_{i}^{*}\tilde{c}_{j}}\langle\psi_{i}^{c}|\hat{b}_{\ell}^{\dagger}\hat{b}_{\ell'}|\psi_{j}^{c}\rangle e^{i\theta(N_{\ell'}^{j}-N_{\ell}^{j})}, & \ell'\geqslant\ell,\\
{\displaystyle \sum_{i,j}\tilde{c}_{i}^{*}\tilde{c}_{j}}\langle\psi_{i}^{c}|\hat{b}_{\ell}^{\dagger}\hat{b}_{\ell'}|\psi_{j}^{c}\rangle e^{i\theta(N_{\ell'}^{j}-N_{\ell}^{j}+1)}, & \ell'<\ell.
\end{cases}
	\label{eq_corr_swap}
\end{equation}
Hence, we obtain that the anyonic correlations, equivalent to Eq.~(\ref{corr_ab_fock}), are formally established in the one-body correlation of the impurity. The spin swapping processes in our model play the role of the operators $\hat{N}_\ell$ in the JW transformation of the AHM. 

\begin{figure}
	\centering\includegraphics[width=1\linewidth]{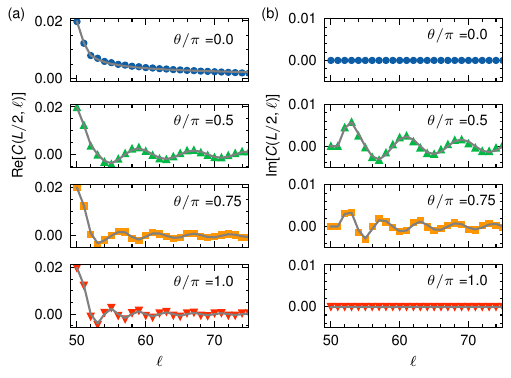} %
	\caption{The real (a) and the imaginary (b) part of the one-body correlation function of the single impurity in our swap model for different $\theta$. We have used $N_\uparrow\!=\!50, N_\downarrow\!=\!1,  L\!=\!100$ and $J_{\text{ex}}\!=\!0.1$. Grey solid lines: results of the AHM with $N\!=\!50, L\!=\!100$ and hard-core interaction. }
	\label{fig_corr}
\end{figure}

This result is illustrated in Fig.~\ref{fig_corr}: the swap model yields identical correlations as the AHM (up to a normalization factor~\cite{supp}). For $\theta=0$, the bosonic correlation shows a power law decay; in the case $\theta=\pi$, we recover the $k_F$-oscillation (with $k_F\!=\!\pi\rho_\uparrow$ being the Fermi momentum and $\rho_\uparrow$ the density of the host particles) due to the fermionic statistics~\cite{2023Basak}. For values of $\theta$ between $0$ and $\pi$, we observe oscillations of period of $2\pi/\theta\rho_\uparrow$ in both the real and the imaginary part of the one-body density correlation, which is a manifestation of the anyonic features therein.

The momentum distribution, derived from the Fourier transform of anyonic correlation functions, serves as a useful observable of anyonic characteristics. In our study, the impurity's quasi-momentum profile mirrors the anyonic behaviour seen in the AHM; see Fig.~\ref{fig1}(c). Crucially, this opens up a new experimental pathway: using an impurity allows one to directly probe anyonic momentum distributions, which are not directly accessible in standard bosonic implementations of AHM~\cite{2011Keilmann,2015Greschner,2016Cardarelli,2016Straeter,2015Tang}.

\paragraph{Effective realization.}
We now demonstrate that the ground state $|\Psi_{\rm swap}(\theta)\rangle$ can be prepared by exciting a spin wave in the swap model at $\theta=0$. This offers a concrete strategy to explore anyonic correlations in two-component Bose gases operating in the strongly-correlated regime. 
Considering a closed spinful chain under the constraint of no double occupancy, spin waves are known to be the low energy excitations~\cite{2020Ivantsov,2022Kesharpu}. They are eigenstates of the spin cyclic permutation operator $\hat{P}$ defined as $\hat{P}|\sigma_{1},\sigma_{2},\cdots,\sigma_{N}\rangle\!=\!|\sigma_{N},\sigma_{1},\cdots,\sigma_{N-1}\rangle$ with $\sigma\!=\!\{\uparrow,\downarrow\}$ and $N \!=\! N_{\uparrow} + N_{\downarrow}$. In the case of a single impurity, the operation of $\hat{P}$ is equivalent to the application of spin swaps. For instance, the spin configuration $|\downarrow\uparrow\uparrow\uparrow\rangle\!=\!\hat{P}|\uparrow\uparrow\uparrow\downarrow\rangle$ can be equivalently obtained from $\hat{\mathcal{E}}_{1,2}\hat{\mathcal{E}}_{2,3}\hat{\mathcal{E}}_{3,4}|\uparrow\uparrow\uparrow\downarrow\rangle$. Thus, for an open chain described by our swap model with $\theta\!=\!0$, where the spin swapping processes are allowed, the spin-wave excitations are expected to exist in the thermodynamic limit. Relating the phase factor $e^{i\theta}$ in our swap model to the eigenvalues of the spin waves $e^{i2\pi p/N}$, with  the spin-wave momenta $p\!=\!0,\cdots,N-1$, the ground states $|\Psi_{\rm swap}(\theta)\rangle$ can be associated to the spin-wave states with different momenta $p$. This intuitive picture not only leads to an alternative way of deriving the anyonic correlation~\cite{supp}, but also indicates that the ground state $|\Psi_{\rm swap}(\theta)\rangle$ can be effectively realized via a slow injection of momentum to the impurity such that the spin-wave states are adiabatically prepared.

To support this intuitive picture, we start from the ground state of our swap model with $\theta\!=\!0$ and then dynamically evolve the system by applying a weak constant force. The time-evolution is governed by the following Hamiltonian, 
\begin{equation}
    \hat{H}_{F}=\hat{H}_{\text{swap}}(\theta=0)-{\displaystyle \sum_{\ell}}Fa\ell\hat{n}_{\downarrow\ell},
    \label{HF}
\end{equation}
where a constant force $F$ is applied on the impurity only and we set the lattice constant $a\!=\!1$ in the following. 
To characterize how close is the time-evolved state $|\Psi_{F}(t)\rangle$ to the ground state $|\Psi_{\rm swap}(\theta)\rangle$, we define the fidelity as
\begin{equation}
    \mathcal{F}(\theta,t)=|\langle\Psi_{\rm swap}(\theta)|\Psi_{F}(t)\rangle|^2.
\end{equation}

\begin{figure}
	\centering\includegraphics[width=0.99\linewidth]{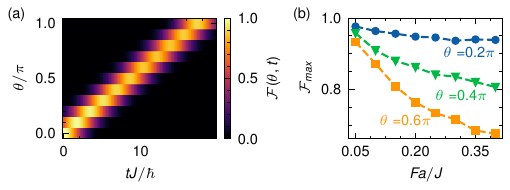} %
	\caption{(a) The fidelity $\mathcal{F}(\theta,t)$ as a function of $\theta$ and evolution time $t$ for $Fa/J=0.1$. (b) The maximum of $\mathcal{F}(\theta,t)$ as a function of force $F$ for different $\theta$. Here we use $N_\downarrow=1, N_\uparrow=10, L=20, J_{\rm ex}=0.1$.} 
	\label{fig_fid}
\end{figure}

In Fig.~\ref{fig_fid}(a), the fidelity $\mathcal{F}(\theta,t)$ is plotted as a function of $\theta$ and $t$. It is observed that the fidelity approaches 1 in proximity to the diagonal, indicating that the ground state $|\Psi_{\rm swap}(\theta)\rangle$ can be effectively approximated by applying a weak force over a specific duration. Taking $\theta/\pi\!=\!\{0.2, 0.4, 0.6\}$ as examples, Fig.~\ref{fig_fid}(b) illustrates the maximum fidelity values as a function of the constant force $F$. It demonstrates that the fidelity converges towards unity as $F$ decreases, which corresponds to a more adiabatic preparation process.

The dynamical evolution governed by the Hamiltonian (\ref{HF}) can be readily implemented in experiments. Due to its $tJ$-type nature, $\hat{H}_{\rm swap}(\theta\!=\!0)$ is a good approximation for the spinful Bose-Hubbard model in the strongly interacting regime~\cite{supp}. In the continuum limit, it can be used to describe the recent experimental observation of anyonization in bosonic systems, where a Bose-Einstein condensate of Cs atoms was loaded into a 2D optical lattice, realizing an array of 1D Bose gases in the Tonks-Girardeau regime~\cite{2024Dhar}. A spin impurity, introduced via a radio-frequency pulse, performed adiabatic motion under a weak force. By varying the evolution time, the system exhibited anyonic features in the impurity's momentum distribution~\cite{2024Dhar}. The lattice nature of our models [Eqs.~(\ref{H_swap}) and (\ref{HF})] can also find other applications, such as tilted optical lattices and a closed chain under rotation. 

\paragraph{Quench dynamics.}

\begin{figure*}
	\centering\includegraphics[width=0.97\linewidth]{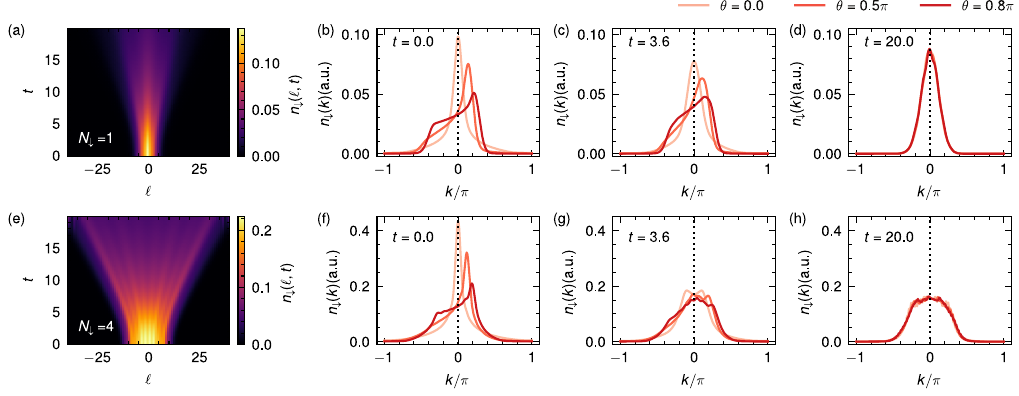}
	\caption{(a) Spatial-temporal density distribution of the single impurity after abruptly releasing the trap for $\theta\!=\!\pi/2$. (b-d) The quasi-momentum distribution of the impurity at different values of expansion time $t$ (in unit of $\hbar/J$). We consider an open chain with $L\!=\!80, N_\downarrow\!=\!1, N_\uparrow\!=\!6, J_{\rm ex}\!=\!0.1$ and $V_{\rm tr}\!=\!0.02$. (e-h) Same as (a-d) but considering $N_\downarrow=4$ impurities.} 
	\label{fig_rapid}
\end{figure*}

It is an interesting perspective to investigate the dynamical properties of the anyonic correlations in our swap model. We start with the ground state of the swap model at a given value of $\theta$ with an overall harmonic trapping potential of strength $V_{\rm tr}$, which exhibits anyonic quasi-momentum distribution for different values of $\theta$; see Fig.~\ref{fig_rapid}(b). At $t\!=\!0$, we abruptly remove the trap and monitor the quasi-momentum distribution of the impurity. Despite distinct initial anyonic momentum distributions for different $\theta$, all curves converge to the same asymptotic distribution after long timescales of expansion; see Fig.~\ref{fig_rapid}(d). 
Such an asymptotic distribution of the impurity reflects its spatial density distribution before removing the external trap~\cite{2021Alam}. Such a rapidity measurement potentially offers a useful tool for probing the \textit{in situ} distribution of a spinor quantum gas.

\paragraph{Multi-impurities.}
So far we have focused our discussion on the case of a single impurity. The anyonic correlations can also be qualitatively engineered in the case of multiple impurities~\cite{supp}. 
As shown in Fig.~\ref{fig_rapid}(f), a very similar asymmetric quasi-momentum distribution can be observed in the system with more impurities (e.g.\ $N_\downarrow\!=\!4$ and $N_\uparrow\!=\!6$). A long-time free expansion leads to a fermion-type quasi-momentum distribution, which is reminiscent of the phenomenon of dynamical fermionization of anyons~\cite{2008Campo}.

\paragraph{Conclusion and outlook.}
We have introduced a minimal 1D swap model that features anyonic correlations, which can be revealed by a simple spin-charge separation analysis. In this framework, spin-swapping processes play a similar role in modifying the one-body correlations as the more conventional Jordan-Wigner string of the anyon-Hubbard model. Moreover, the ground state of the swap model can be prepared to good approximation via a straightforward protocol, offering a realistic route to engineer anyonic behaviour in quantum engineered systems. As an interesting perspective, one can envision implementing the complex phase in the swapping interaction by designing a Floquet-driven protocol~\cite{2017Eckardt}. Generalizing our scheme to non-Abelian anyons~\cite{2001Kitaev} --- or to other exotic particles such as traid anyons~\cite{2020Harshman,2022Harshman,2024Nagies} and paraparticles~\cite{2023Nielsen,Wang2025} encoded in the impurity dynamics --- remains an exciting open question.

\begin{acknowledgments}
    We thank Xi-Wen Guan, Felix A. Palm and Yunbo Zhang for helpful discussions.
	Work in Brussels is supported by the ERC (LATIS project), the EOS project CHEQS, the FRS-FNRS Belgium and the Fondation ULB. The Innsbruck team acknowledges funding by a Wittgenstein prize grant with the Austrian Science Fund (FWF) project number Z336-N36, by the European Research Council (ERC) with project number 789017, by an FFG infrastructure grant with project number FO999896041, and by the FWF's COE 1 and quantA. Y.G. is supported by the FWF with project number 10.55776/COE1. 
    Computational resources have been provided by the Consortium des Équipements de Calcul Intensif (CÉCI), funded by the Fonds de la Recherche Scientifique de Belgique (F.R.S.-FNRS) under Grant No. 2.5020.11 and by the Walloon Region. Numerical simulations are based on matrix product states implemented using ITensors~\cite{2022_itensor,2022_itensor03}.

    B.W. and A.V. contributed equally to this work.
\end{acknowledgments}

\bibliography{mybib}

\clearpage
\appendix

\onecolumngrid
\setcounter{equation}{0}
\setcounter{figure}{0}
\renewcommand{\theequation}{S\arabic{equation}}
\renewcommand{\thefigure}{S\arabic{figure}}
\renewcommand{\thesection}{S\arabic{section}}
\renewcommand{\thesubsection}{\thesection.\arabic{subsection}}
\renewcommand{\thesubsubsection}{\thesubsection.\arabic{subsubsection}}

\begin{center}
	{\Large\bfseries Supplementary Materials}
\end{center}

\section{I.~~ Variations in the swap model}\label{sec_A}
\begin{figure}[H]
	\centering\includegraphics[width=0.9\linewidth]{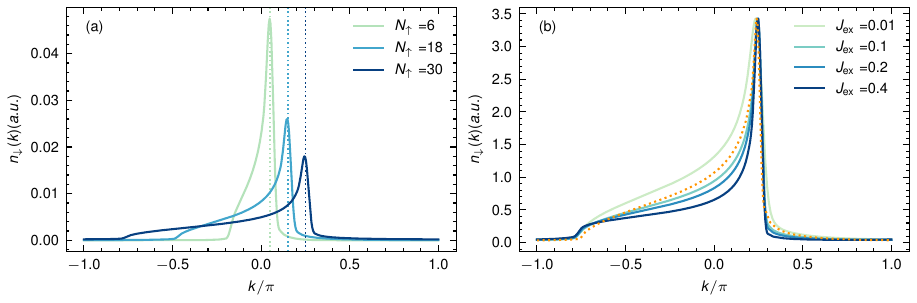} %
	\caption{(a) Quasi-momentum distribution of impurity for different number of host particles $N_{\uparrow}$ with $J_\text{ex}=0.1, \theta=\pi/2, L=60$.  The dotted lines indicate the location of the peak which is associated with $k_{\text{peak}}=\theta N_{\uparrow}/L$.  (b) Quasi-momentum distribution of impurity for different $J_{\text{ex}}$ with $\theta=\pi/2, N_{\uparrow}=30, L=60$. The dotted orange line represent the results of the AHM with hard-core interaction and $N=30, L=60$. The peak has been rescaled for a better illustration.} 
	\label{fig_Jex}
\end{figure}

The similarity between Eqs.~(\ref{corr_ab_fock}) and (\ref{eq_corr_swap}) in main text suggests that the many-body correlation of anyons (\ref{corr_ab_fock}) is encoded in the single particle density matrix of the impurity. The impurity helps to form a many-body correlated state along with the host particles. As shown in Fig.~\ref{fig_Jex}(a), the peak of the anyonized quasi-momentum distribution is shifted by the number of host particles. The peak location is found to be $k_{\text{peak}}=\theta N_{\uparrow}/L$, which share a similar form of the Fermi momentum $k_F=\pi N/L$ in the case of a 1D Fermi gas on a lattice. 

As our swap model is intrinsically different from the anyon-Hubbard Model (AHM), a quantitative agreement between their quasi-momentum distributions is not straightforward. We treat $J_{\rm ex}$ as a free parameter in our model. As depicted in  Fig.~\ref{fig_Jex}(b), the values of $J_{\rm ex}\approx0.1J$ match the results obtained from the AHM. Smaller values of $J_{\rm ex}$ seem to broaden the quasi-momentum distribution, so does a smaller system size (not shown). This can be understood intuitively: smaller $J_{\rm ex}$ would effectively reduce the space that an impurity could explore/swap, thus leading to a larger uncertainty in momentum space.

Note that by means of a gauge transformation $\hat{b}_{\downarrow\ell} \rightarrow e^{i\theta \ell}\hat{b}_{\downarrow\ell}$, the above Hamiltonian can be rewritten as,
\begin{align}
	\hat{H}_{\text{swap}}=-J\sum_{\ell=1}^{L-1}\hat{b}_{\uparrow\ell}^{\dagger}\hat{b}_{\uparrow\ell+1}-Je^{i\theta}\sum_{\ell=1}^{L-1}\hat{b}_{\downarrow\ell}^{\dagger}\hat{b}_{\downarrow\ell+1}-J_{\text{ex}}\sum_{\ell=1}^{L-1}\hat{b}_{\uparrow\ell}^{\dagger}\hat{b}_{\downarrow\ell+1}^{\dagger}\hat{b}_{\downarrow\ell}\hat{b}_{\uparrow\ell+1}+\mathrm{h.c.}
	\label{H_Jimp}
\end{align}
This Hamiltonian has a simple interpretation:~considering periodic boundary conditions, this Hamiltonian describes hardcore particles moving on a ring, where one species ($\downarrow$) feels a magnetic flux $\theta$, while the other ($\uparrow$) does not; and these two components are coupled through nearest-neighbor spin swapping processes. These basic ingredients lead to signatures of anyonic correlations in the one body correlation function.

\section{II.~~ Derivation in terms of spin waves}
In this section, we show that in the case of a \textit{single} impurity, the one-body correlation of the single impurity can be derived by considering spin-wave states in $\hat{H}_{\rm swap}(\theta=0)$, the swap model with vanishing $\theta$. For convenience, we consider a closed chain. Due to the constraint of no double occupancy, we can apply a spin-charge separation treatment~\cite{2020Ivantsov,2022Kesharpu,2023Basak} and write an eigenstate of $\hat{H}_{\rm swap}(\theta=0)$ as $|\Psi\rangle=|\varphi\rangle\otimes|\chi\rangle$, where $|\varphi\rangle$ and $|\chi\rangle$ represents the charge and the spin part, respectively. Because of the spin swapping processes, which are equivalent to the spin cyclic permutation, the low energy excited states are spin waves~\cite{2020Ivantsov,2022Kesharpu},
\begin{equation}
    |\chi\rangle=\frac{1}{\sqrt{N}}\sum_{s=0}^{N-1}e^{i\tilde{\theta} s}|\chi_{s}\rangle.
    \label{eq_spin_wave}
\end{equation}
Here, we associate the spin-wave momentum $p=0,1,\cdots,N-1$ to the angle $\tilde{\theta}=2\pi p/N$. The spin sector are expanded in the spin configuration basis $\{|\chi_{s}\rangle\}$ with $|\chi_{s}\rangle=|\sigma_{1},\cdots,\sigma_{\ell_{p}},\cdots,\sigma_{N}\rangle,\sigma_{\ell_{p}}=\uparrow,\downarrow$ and $s$ labels the position of the single spin-down. For example, in the case of one spin-up and one spin-down, we have
\begin{equation}
    |\chi(\tilde{\theta}=0)\rangle=\frac{1}{\sqrt{2}}\left(|\uparrow\downarrow\rangle+|\uparrow\downarrow\rangle\right),~~|\chi(\tilde{\theta}=\pi)\rangle=\frac{1}{\sqrt{2}}\left(|\uparrow\downarrow\rangle-|\uparrow\downarrow\rangle\right).
\end{equation}

\begin{figure}[H]
	\centering\includegraphics[width=0.5\linewidth]{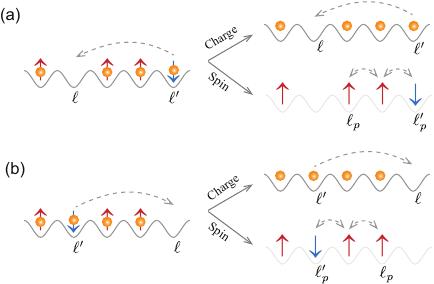} %
	\caption{ (a) Sketch of the one-body correlator of the impurity in the case of $\ell'>\ell$. The swapping processes start with the fourth spin $\ell'_p=N_\ell'+1=4$ and end with the second spin $\ell_p=N_\ell+1=2$. (b) Sketch of the one-body correlator of the impurity in the case of $\ell'<\ell$. The swapping processes start with the fourth spin $\ell'_p=N_\ell'+1=2$ and end with the second spin $\ell_p=N_\ell=4$. Notice the difference between $\ell_p=N_\ell+1$ in (a) and $\ell_p=N_\ell$ in (b). } 
	\label{fig_scs}
\end{figure}

Our interest is to compute the one-body correlator of the impurity, which can also be expressed in a spin-charge separated form~\cite{2023Basak}. For $\ell'>\ell$, we have,
\begin{equation}
    \langle\hat{b}_{\downarrow\ell}^{\dagger}\hat{b}_{\downarrow\ell'}\rangle=\langle\varphi|\hat{b}_{\ell}^{\dagger}\hat{b}_{\ell'}\delta_{\hat{N}_{\ell}}^{\ell_{p}-1}\delta_{\hat{N}_{\ell'}}^{\ell_{p}'-1}|\varphi\rangle\cdot\langle\chi|\hat{\mathcal{E}}_{\ell_{p},\ell_{p}+1}\cdots\hat{\mathcal{E}}_{\ell'_{p}-1,\ell'_{p}}|\chi\rangle.
    \label{eq_corr_imp}
\end{equation}
which is composed of the hopping in the charge sector and swapping in the spin sector. As illustrated in Fig.~\ref{fig_scs}, the Kronecker delta ensures that the swapping processes start and end with the right spins. 

For the low energy excited states that host spin waves, the spin part in Eq.~(\ref{eq_corr_imp}) becomes
\begin{equation}
    \langle\chi|\hat{\mathcal{E}}_{\ell_{p},\ell_{p}+1}\cdots\hat{\mathcal{E}}_{\ell'_{p}-1,\ell'_{p}}|\chi\rangle=\frac{1}{N}e^{i\tilde{\theta}(\ell'_{p}-\ell_{p})}.
    \label{eq_spin}
\end{equation}
Combining Eqs.(\ref{eq_corr_imp}) and (\ref{eq_spin}), the one-body correlation of the impurity can be written as
\begin{equation}
    \langle\hat{b}_{\downarrow\ell}^{\dagger}\hat{b}_{\downarrow\ell'}\rangle|_{\ell'>\ell}=\frac{1}{N}\langle\varphi|\hat{b}_{\ell}^{\dagger}\hat{b}_{\ell'}e^{i\tilde{\theta}(\hat{N}_{\ell'}-\hat{N}_{\ell})}|\varphi\rangle,
\end{equation}
and similarly, for $\ell'<\ell$, we have
\begin{equation}
    \langle\hat{b}_{\downarrow\ell}^{\dagger}\hat{b}_{\downarrow\ell'}\rangle_{\ell'<\ell}=\frac{1}{N}\langle\varphi|\hat{b}_{\ell}^{\dagger}\hat{b}_{\ell'}e^{-i\tilde{\theta}(\hat{N}_{\ell}-\hat{N}_{\ell'}-1)}|\varphi\rangle.
\end{equation}
Thus the one-body correlator of the impurity is equivalent to the one-body correlator of anyons (upto a normalization factor). For the purpose of calculating the impurity correlator, choosing a particluar spin-wave excited state with a finite value of $\tilde{\theta}$ (i.e. limiting to that spin-wave sector) is similar to considering the ground state of the swap Hamiltonian $\hat{H}_{\text{swap}}$ with a finite swapping phase ($\theta=\tilde{\theta}$). 

\section{III.~~ Time evolution governed by $\hat{H}_{\rm sBHM}$}
The spinful Bose-Hubbard model (sBHM) in the strongly-interacting regime takes the following form
\begin{align}
	\hat{H}_{\text{sBHM}}= -J\sum_{\ell= 1}^{L-1}\left(\hat{b}_{\uparrow\ell}^{\dagger}\hat{b}_{\uparrow\ell+1}+\hat{b}_{\downarrow\ell}^{\dagger}\hat{b}_{\downarrow\ell+1}+h.c.\right)+U_{\uparrow\downarrow}\sum_{\ell}\hat{n}_{\uparrow\ell}\hat{n}_{\downarrow\ell}-\sum_{\ell}Fa\ell\hat{n}_{\downarrow\ell}.
	\label{H_BHM}
\end{align}
Here we consider intra-component interaction in the hard-core limit, i.e. $U_{\uparrow\uparrow}\rightarrow\infty$ and $U_{\downarrow\downarrow}\rightarrow\infty$. The inter-interaction between host particles and the impurity is denoted by $U_{\uparrow\downarrow}$. A constant force $F$ is applied only for the impurity. In the strongly-interacting regime, the on-site interaction terms contribute to a high-energy subspace. Thus considering the configurations with no doubly-occupied sites, the Hamiltonian can be effectively described by a low-energy Hamiltonian.  The $tJ$-model is one of such effective low-energy models, where the exchange terms and nearest-neighbor interaction terms replace the on-site interaction terms in the Hubbard model~\cite{2019Eckle}. Similarly, the Hamiltonian $\hat{H}_F$ [see Eq.~\eqref{HF} in main text] (with the nearest-neighbor interaction terms being neglected) is also expected to approximate the sBHM in the storngly-interacting regime.

\begin{figure}[H]
	\centering\includegraphics[width=0.5\linewidth]{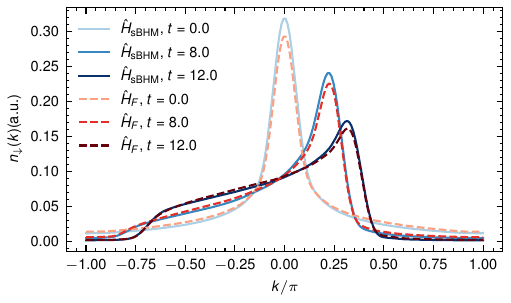} %
	\caption{ Quasi-momentum distribution of impurity after different evolution time. Solid lines: obtained from the evolution governed by the spin Bose-Hubbard model $\hat{H}_{\rm sBHM}$ with $U_{\uparrow\downarrow}\!=\!40$. Dashed lines: the results correspond to the effective Hamiltonian $\hat{H}_F$ with $J_{\rm ex}\!=\!0.1$. In both cases we have considered $N_\downarrow\!=\!1, N_{\uparrow}\!=\!10, L\!=\!20$ and $F\!=\!0.1$, so that the different evolution time $t\!=\!0, 8.0, 12.0$ corresponds to $\theta\!=\!Ft/\rho_\uparrow\approx0, \pi/2, 3\pi/4$, respectively. We set $J=1$ (resp. $\hbar/J=1$) as the unit of energy (resp. time) and lattice constant $a=1$.} 
	\label{fig_BHM}
\end{figure}

Taking the ground state of $\hat{H}_{\rm sBHM}(F=0)$  as the initial state, we simulate the quench dynamics by solving the time dependent Schr\"odinger equation associated with $\hat{H}_{\rm sBHM}$ with a weak force $Fa=0.1J$. In Fig.~\ref{fig_BHM}, we plot the quasi-momentum distributions of the impurity at different time, that exhibit qualitatively a good agreement with the results obtained from the dynamical evolution governed by $\hat{H}_F$. 
It suggests an efficient and practical way of realizing anyonic correlations in a strongly-interacting 1D quantum gas.

\section{IV.~~ Multiple impurities}
\begin{figure}[H]
	\centering\includegraphics[width=0.5\linewidth]{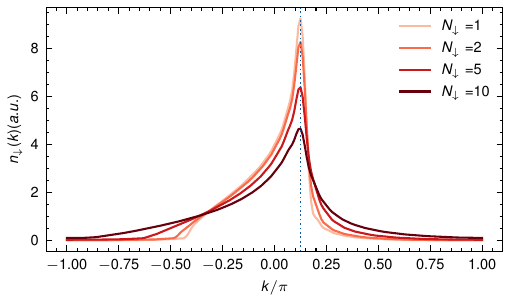} %
	\caption{Quasi-momentum distribution of different number of impurities in the swap model. We consider $L=40, N_{\uparrow}=10, J_\text{ex}=0.1$ and $\theta=\pi/2$. The dotted line indicates the location of the peak which is associated with $k_{\text{peak}}=\theta N_{\uparrow}/L$.} 
	\label{fig_Nimp}
\end{figure}

In the swap model, we find that the anyonization features, i.e.\ the asymmetric quasi-momentum distribution, survive in the case of multiple impurities. As depicted in Fig.~\ref{fig_Nimp}, a qualitative agreement is found for different number of impurities. Because the swapping phases are involved only with the swapping between different species, in the expression of one-body correlation of the impurities, the number of charges $N_\ell$ in the phases factors are replaced by the number of host particles $N_{\uparrow\ell}$, i.e. 
\begin{equation}
    \langle\hat{b}_{\downarrow\ell}^{\dagger}\hat{b}_{\downarrow\ell'}\rangle={\displaystyle \sum_{i,j}\tilde{c}_{i}^{*}\tilde{c}_{j}}\langle\varphi_{i}^{c}|\hat{b}_{\ell}^{\dagger}\hat{b}_{\ell'}|\varphi_{j}^{c}\rangle e^{i\theta(N_{\uparrow\ell'}^{j}-N_{\uparrow\ell}^{j})}.
\end{equation}
After a Fourier transformation, the location of the peak in the quasi-momentum distribution is thus determined by the number of host particles, i.e. $k_{\text{peak}}=\theta N_{\uparrow}/L$, as indicated by the dotted line in Fig.~\ref{fig_Nimp}.

\end{document}